# The audiovisual resourse as a pedagogical tools in times of covid 19. An empirical analysis of its efficiency

Juan Rodriguez Bassignana y Carolina Asuaga

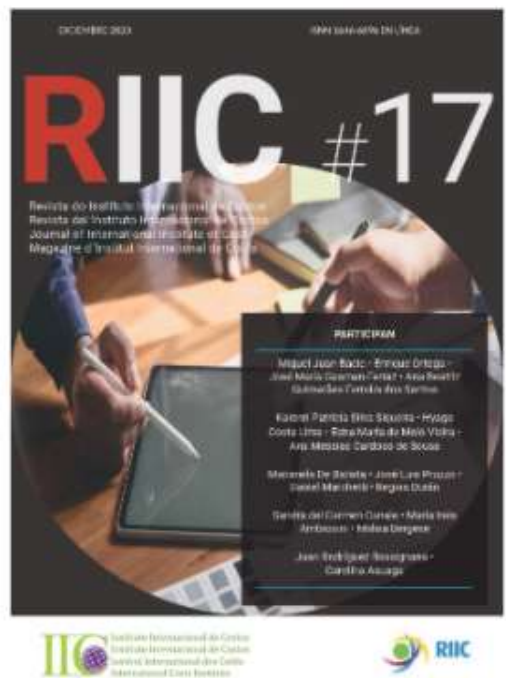

JEL: A20, A22

**ABSTRACT**

The global pandemic caused by the COVID virus led universities to a change in the way they teach classes, moving to a distance mode. The subject "Modelos y Sistemas de Costos " of the CPA career of the Faculty of Economic Sciences and Administration of the Universidad de la República (Uruguay) incorporated audiovisual material as a pedagogical resource consisting of videos recorded by a group of well experienced and highest ranked teachers.

The objective of this research is to analyze the efficiency of the audiovisual resources used in the course, seeking to answer whether the visualizations of said materials follow certain patterns of behavior. 13 videos were analyzed, which had 16,340 views, coming from at least 1,486 viewers.

It was obtained that the visualizations depend on the proximity to the test dates and that although the visualization time has a curve that accompanies the duration of the videos, it is limited and the average number of visualizations is 10 minutes and 4 seconds. It is also concluded that the efficiency in viewing time increases in short videos.

**RESUMEN**

La pandemia mundial provocada por el virus COVID llevó a las universidades a un cambio en la forma de impartir las clases pasando a una modalidad a distancia. La materia Modelos y Sistemas de Costos de la carrera de contador público de la Facultad de Ciencias Económicas y de Administración de la Universidad de la República (Uruguay) incorporó como recurso pedagógico material audiovisual consistente en videos grabados por los docentes de mayor rango. El objetivo de esta investigación es analizan la eficiencia de los recursos audiovisuales utilizados en el curso, buscando dar respuesta a si las visualizaciones de dichos materiales siguen ciertos patrones de comportamiento. Se analizaron 13 videos, los cuales tuvieron 16.340 visualizaciones, provenientes de al menos 1486 espectadores. Se obtuvo que las visualizaciones dependen de la cercanía a las fechas de las pruebas y que aunque el tiempo de visualización tiene una curva que acompaña a la de la duración de los videos, la misma está acotada y el promedio de visualizaciones es de 10 minutos 4 segundos. Se concluye también que la eficiencia en el tiempo de visualización aumenta en los videos de corta duración.

## 1) INTRODUCCIÓN

El virus COVID 19 ha traído consecuencias que van más allá de lo sanitario, impactado fuertemente en la economía y propiciando un cambio en prácticamente todas las actividades humanas, entre ellas la enseñanza.

El cierre forzoso de los espacios físicos de las universidades llevó a los docentes a recurrir a modalidades de enseñanza virtuales sin el conocimiento necesario para ello y sin tiempo suficiente para un correcto aprendizaje de las características y requisitos necesarios que esta modalidad de enseñanza requiere.

En Uruguay, la Universidad de la República -principal institución de educación terciaria, gratuita y pública- cierra el acceso a sus instalaciones el 15 de marzo de 2020, a dos días de detectado el primer caso de COVID 19 en el País, pasando a dictar las clases en formato virtual. Al finalizar el semestre, se dictaron el 96% de las asignaturas de las 184 carreras que se ofrecen. Se desarrollaron 2.720 cursos en formato virtual. Se dictaron más de 24.000 clases para más de 100.000 alumnos. El 92% de los estudiantes terminó al menos un curso en el semestre.[1]

La Facultad de Ciencias Económicas y de Administración había comenzado el dictado del semestre apenas una semana antes de decretarse la emergencia sanitaria, y luego de una pausa de entre 6 y 12 días, se comenzaron a dictar la totalidad de sus cursos en forma virtual.

---

[1] Fuente: Material preparado por la Asociación de Docentes de la Universidad de la República (ADUR).

La materia Modelos y Sistemas de Costos, de la carrera de contador público, atendió de forma virtual a más de 1.000 estudiantes, mediante 6 grupos teórico-prácticos. El objetivo de esta investigación es analizan la eficiencia de los recursos audiovisuales utilizados en el curso, buscando dar respuesta a si las visualizaciones de dichos materiales siguen ciertos patrones de comportamiento y la forma en que el estudiante interactúa con ellos.

## 2) MARCO TEORICO

Es complejo definir el comienzo de lo que hoy se conoce como educación a distancia. Algunos autores datan su origen en las cartas instructivas de la civilización sumeria, destacándose también las cartas formativas de la Grecia Antigua; y fundamentalmente en la civilización romana con los aportes de Cicerón, Horacio y, especialmente Séneca, autor de más de un centenar de cartas que constituyen en su conjunto una unidad didáctica de filosofía (Alfonso, 2003; Arazomena,1992). La aparición de la imprenta y la popularización del correo postal facilitó la circulación de materiales de estudio, pero existe consenso en la literatura que la educación a distancia en la concepción del concepto dado por la UNESCO (2009)[2], tiene sus orígenes en 1792, cuando Caleb Phillips, anuncia en la Gaceta de Boston un curso a distancia de caligrafía, ofreciendo la posibilidad de tutorías por correspondencia.

Orta y Barón (2012) en un cuidadoso análisis temporal, datan en 1889 el primer curso universitario a distancia, ofrecido por la Queen´s University of Kingstone, en Canadá. Unos años más tarde, también en Canadá, la Universidad de Saskatckewan ofrece la posibilidad de formación en determinados programas sin acudir a clase. En la década del 30 se crea en Francia el Centro Nacional de Enseñanza a Distancia, y en 1945 nace en Sudáfrica la UNISA, primera universidad de educación a distancia.

Pero el primer caso de práctica masiva de esta modalidad surge en 1969 con la aparición de la Open University del Reino Unido, universidad gratuita y de financiación pública. 4 años después se crea en España la UNED, la cual también es financiada e impulsada por el gobierno, comenzando luego a surgir paulatinamente en diversas partes del mundo universidades con modalidad a distancia, tanto públicas como privadas.

---

[2] Entendida como el uso de técnicas pedagógicas, recursos y medios de comunicación para posibilitar el proceso de aprendizaje cuando alumnos y docentes se encuentran separados en el espacio y/o tiempo.

Si bien este tipo de universidades han logrado un amplio desarrollo, Villalonga (2016) menciona que las mismas aún no han sorteado dos grandes dificultades: la alta tasa de deserción y el aún poco reconocimiento a nivel académico y laboral.

En paralelo al desarrollo institucional de la educación a distancia, en los últimos 30 años han existido significativos avances en el ámbito de la tecnología de la información y las comunicaciones (TIC), que dio lugar a un incremento exponencial en el uso y desarrollo de las computadoras personales, las *tablets* y los *smartphones*, los cuales en conjunto con el aumento creciente a nivel mundial de la accesibilidad a internet y el desarrollo de la nube de datos, ayudaron a que se desarrollara y popularizara el *e-learning*. Cabe destacar que éste se basa en un conjunto de múltiples herramientas pedagógicas virtuales, como por ejemplo los entornos virtuales de aprendizaje, los canales de *streaming* donde se divulga contenido educativo, las *apps* educativas, las videoconferencias masivas, la gamificación educativa o las bibliotecas virtuales.

Cabe señalar, que las TIC posibilitaron el acercamiento de los contenidos educativos a través de diferentes modalidades como los *Opening Course Wares*, donde un consorcio de universidades comparten el contenido dictado en sus clases a iniciativa del Instituto Tecnológico de Massachusetts (MIT), el que comenzó esta práctica con sus propios cursos en el año 2001, dando lugar a los *Massive Online Opening Courses* o MOOC, los cuales han sido masificados, en especial desde el año 2012 con el lanzamiento de *Coursera* y *Udacity* (en ambos casos por docentes de la Universidad de Stanford) y en el mismo año, el MIT y la Universidad de Harvard en conjunto e institucionalmente, presentaron la plataforma EDX.

Los MOOC ofrecidos por estas plataformas son gratuitos (*freemiun),* únicamente se paga un monto si se quiere obtener un título una vez finalizado exitosamente; o pueden ser de bajo costo y no exigir los requisitos de entrada usuales para habilitar el cursado de una carrera universitaria (secundaria terminada, mayoría de edad o examen de ingreso). Si bien en un inicio los MOOC se desarrollaron en Estados Unidos, paulatinamente surgen cursos a distancia en diversas universidades del mundo. Por ejemplo y desde la disciplina de costos, en junio de 2019 la Universidad de Chile lanzó en Coursera un MOOC de Costos para los Negocios.

El desarrollo de esta modalidad ha permitido apreciar a nivel académico la calidad que puede alcanzar este recurso educativo y varias universidades han aceptado, en periodo pre pandemia, que ciertos cursos de MOOC sean tomados como créditos de carreras, y se observababa una tendencia de crecimiento en el número de cursos de grado o postgrado en modalidad semipresencial o en algunos casos directamente a distancia.

Este mix entre educación a distancia y TIC ha generado una nueva forma de aprendizaje denominada "*Blended Learning*", donde se combina el aula presencial con las herramientas *online* e interacciones virtuales entre docente y alumnos. Dentro de esta forma de aprendizaje, la más difundida es la del aula invertida o *Flipped Classroom*, donde la tarea del estudiante se centra en reproducir videos, escuchar *podcast* o leer material disponible en plataformas virtuales, buscando que la clase presencial se dedique básicamente a practicar en grupo y colaborativamente, propiciando una alta participación del alumno y fomentando el trabajo en equipo, cambiando el rol del docente de dueño y trasmisor del conocimiento a facilitador del mismo[3]. Por motivos de espacio no se desarrollarán los conceptos de *flipped classrom*. Si el lector quiere profundizar en el tema, entre los principales aportes se destacan los de Tucker, (2012); Herried y Schiller,(2103); Bishop y Verleger, (2013); Gilbot et al, (2015); Akçayır y Akçayır,(2018); Hew y Lo, 2018;. Awidi, y Paynter, (2019). Sobre el efecto COVID 19 en la educación terciaria, ya existen estudios científicos que tratan las nuevas técnicas de aprendizaje en el nuevo contexto mundial, en especial la *flipped classroom*, como los de Chick et al (2020); Naresh, (2020); Pitt et al (2020), Strelan (2020), entre otros.

---

[3] Esta metodología comenzó en 2007 cuando Bergmann y Sams, docentes de química, ante la constante ausencia de parte del público estudiantil debido a la distancia de sus hogares con el centro de estudio, realizaron videos donde presentaban Powerpoint explicadas con grabaciones de voz. La masividad de esta técnica vino dada con la aparición de Khan Academy en ese mismo año en Estados Unidos. Si bien la metodología fue concebida originalmente para la educación secundaria, la misma gradualmente se ha aplicado también en el ámbito universitario.

A efectos de determinar los requerimientos para el desarrollo de un curso a distancia o en modalidad *blended* se destacan:

### 2.1 Conocimiento docente

Dominar las tecnologías de la información y manejarlas adecuadamente, es crítico para poder realizar exitosamente un curso a distancia o en formato *blended learning* (Yang, 2015). Surge el concepto de "Conocimiento Tecnológico y Pedagógico del Contenido", conocido como TPACK por sus siglas en inglés [4] el que se basa y profundiza la construcción de Lee Shulman en la década del 80 sobre Contenidos del Conocimiento Pedagógico (PCK)[5] de forma de incluir conocimientos tecnológicos.

La concepción de TPACK se ha desarrollado a partir de los aportes de Mishra y Koehlen (Veáse Mishra y Koehler 2006, Koehler y Mishra 2009, Koehler et al 2015). Los autores analizan el vínculo entre contenido, pedagogía y tecnología, además de las relaciones entre estos mismos y entre todos ellos, definiendo:

*Conocimiento sobre el contenido (CK)***:** es el saber que el docente ha construido sobre la disciplina que enseña y es de importancia crítica.

*Contenido Pedagógico (PK):* es el conocimiento profundo que tienen los docentes sobre los procesos y prácticas o métodos de enseñanza y aprendizaje

*Conocimiento sobre la Tecnología (TK):* Se centra en el conocimiento sobre el uso de recursos tecnológicos, así como la capacidad de reconocer cuando y cuáles herramientas tecnológicas

---

[4] *Technology, pedagogy, and content knowledge*
[5] Para profundizar en PCK véase Schulman 1986 y 1988, Grossman et al, 2011; Cochran et al, 1993,

facilitan la consecución de un objetivo, debiendo el docente adaptarse continuamente a los cambios tecnológicos.

La interacción e intercesión entre estos conocimientos se muestran en la siguiente figura:

Figura 1: El Modelo TPACK

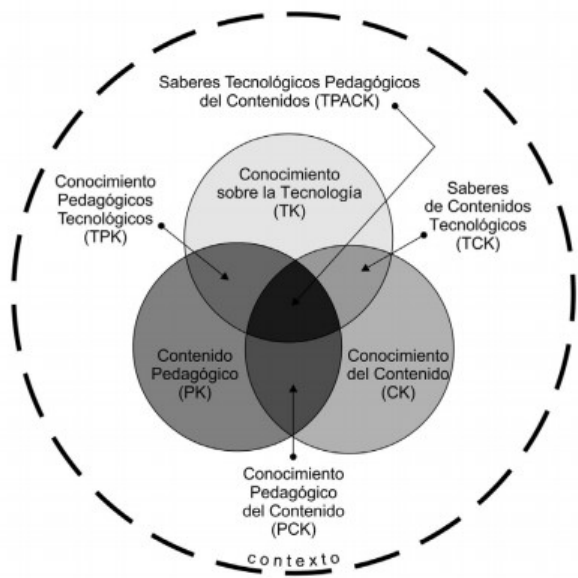

Fuente: Koehler et al 2015

Este modelo lleva a que el docente debe, además de dominar el contenido del curso y las estrategias de enseñanza-aprendizaje, conocer las posibles herramientas tecnológicas existentes y posibles de utilizar, y cómo aplicarlas. Cabe señalar que el estado óptimo de conocimiento se ubica en la intersección de los tres círculos principales. Asimismo, el círculo exterior representa el contexto y tiene como objetivo enfatizar el hecho de que la tecnología, pedagogía y contenidos están inmersos en contextos específicos de enseñanza y aprendizaje. En otro orden, no sólo el dictado del curso debe tener un abordaje diferente en una modalidad a distancia, sino que el docente debe considerar la complejidad que implican las evaluaciones en modalidad virtual.

Asuaga et al (2012) señalan las ventajas de las pruebas en línea al reducir el tiempo de la prueba, dando resultados instantáneos y fáciles de administrar, pero sostienen que la evidencia empírica muestra que la implementación de los sistemas de evaluación basados en la informática, también llamados CBA (Computer Based Assessment systems) ha sido dispar y no han tenido una acogida masiva. Sostienen que la discusión académica se centró en el estrés que podría provocarle al alumno un cambio en el formato de las pruebas, motivada en parte por la aparición en 2004 de los resultados de una investigación base a 1474 estudiantes, quienes manifestaron temores ante posibles problemas técnicos con el equipo, problemas de concentración, y el agregar un estrés adicional al examen. El éxito de un CBA depende de la aceptación por los estudiantes (Terzis y Economides ,2011). Ante esto Toki y Caukill (2003) señalan la importancia de que el estudiante esté familiarizado con el formato, el que puede darse a conocer por medio de simuladores, y que el éxito dependerá de los conocimientos y habilidades de los docentes.

**2.2 Recursos audiovisuales**

Entre los diversos recursos audiovisuales es posible destacar:

*Clases Presenciales Grabadas:* Implican una grabación de audio de un docente dictando una clase frente a un grupo de alumnos. A efecto didáctico, un aspecto negativo es su larga duración. Una posible solución es que se pueda editarse la clase, acortando su duración a un plazo máximo de 20 minutos. Como ventaja se señala que es una herramienta de bajo costo enfocada a clases de tipo práctico, siendo prudencial renovar la oferta de videos cada 2 o 3 años aún incluso cuando los cambios curriculares relevantes sucedan en un plazo mayor.

*Video Conferencias:* Esta herramienta posibilita la sustitución de las clases tradicionales, donde se reemplaza un aula presencial con una conexión vía streaming en tiempo real. Cabe destacar que no es una novedad pedagógica[6] sino tecnológica.

*Audiovisuales:* realización de videos grabados donde se incluyen gráficas, animaciones, links, cuestionarios, entre otras. Aunque la duración de los videos varía, se recomiendan videos cortos (Sexton, 2006; de la Fuente et al 2003; Bell y Bell, 2010) y cabe destacar el formato 5ELFA, tal como lo presentan Asiksoy y Ozdmli (2017), donde la duración máxima de los videos es de 3 minutos para explicar el concepto, 2 minutos para profundizar o ejemplificar el tema expuesto, y 14 minutos para explicar en profundidad o presentar un caso práctico.

### 2.2 Plataformas educativas

Con el devenir del siglo XXI, ha surgido un creciente interés por las plataformas educativas o LMS[7] que ofrecen al alumno y docente diferentes servicios que van desde el alojamiento y repositorios de documentación y recursos audiovisuales, comunicaciones por medio de foros y correo electrónico, hasta la posibilidad de la realización de evaluaciones acreditativas, diagnósticas o formativas, en un ambiente cerrado y controlado. Cabe destacar que actualmente, las plataformas también ofrecen comunicación sincrónica.

No es objeto de este artículo describir el funcionamiento de las plataformas educativas, sino remarcar la importancia de la existencia de estas a efectos de desarrollar tanto un curso a distancia como en modalidad *blended.* A raíz de la emergencia sanitaria declarada a raíz del virus Convid 19, las plataformas educativas fueron cruciales para poder continuar con los

---

[6] Puede llegar a ser una novedad pedagógica, si por ejemplo se utiliza una herramienta como un foro de intercambio.

[7] Learning management systems

procesos de enseñanza aprendizaje, aún cuando dichos entornos hubiesen sido subutilizados.[8]

### 3) PRESENTACION DEL CASO Y METODOLOGIA

La Unidad Académica de Costos y Control de Gestión, del Departamento de Contabilidad y Tributaria de la Facultad de Ciencias Económicas y de Administración de la Universidad de la República (Uruguay), dicta 6 materias semestrales, entre las que se encuentra Modelos y Sistemas de Costos, materia obligatoria de la carrera de Contador Público, y que forma parte de la currícula del quinto semestre (sobre un total de 8 semestres) y aporta 10 créditos de los 360 necesarios para el egreso.

Hasta el año 2019 inclusive, tenía un método de evaluación acreditativa que constaba de dos pruebas parciales de múltiple opción (POM).

La población estudiantil que inicia el cursado de la materia se ubica en un entorno de 1.000 a 1.200 estudiantes, y las clases son dictadas en aulas con capacidad de 80 a 200 alumnos, en salas tradicionales donde el docente da la clase frente a los alumnos contando con dos pizarras de marcador y conexión a proyector, siendo la conexión wifi no del todo funcional.

Respecto a los recursos que disponían los estudiantes en la plataforma que brinda la Universidad (Moodle), contaban con el material teórico escrito, ejercicios prácticos basados en simulaciones abstractas de la realidad, donde el estudiante se enfrenta a situaciones en donde debe utilizar de forma práctica los conocimientos teóricos dados. También los estudiantes tenían a su disposición pruebas parciales y exámenes anteriores.

---

[8] Si el lector quiere profundizar en torno al papel de las plataformas educativas en relación a la pandemia reciente, puede ver Munin &Hasan 2020; Kalcu et al 2020; Gunawan et al, 2020; Ortega, 2020; Sáiz-Manzanares et al, 2020; Young & Donovan, 2020, entre otros.

Aún no se había generado material audiovisual propio de la materia (si existía para otras materias de la Unidad Académica), y los recursos que utilizaban los docentes eran básicamente planillas electrónicas y power point o prezi de elaboración propia.

Tal como fue expuesto en la introducción, a la semana de inicio de las clases la alerta sanitaria provocada por el COVID 19 llevó a discontinuar las clases presenciales, debiendo realizarse el resto del curso en modalidad a distancia.

Respecto a las clases, se hizo énfasis en la parte práctica, y la modalidad elegida para su dictado fue la de videoconferencias mediante la plataforma Zoom, luego de tener una experiencia fallida con la plataforma Big Blue Button.

Un cambio sustancial fue lo realizado con los contenidos teóricos del curso, donde como respaldo a los docentes a cargo de los cursos, los docentes de rango más alto se encargaron de realizar videos explicativos de las 6 unidades teóricas del curso. El método de grabación fue vía Big Blue Button y Zoom, luego de lo cual se editaban los videos mediante las herramientas Camstacia y ShotCut, para luego subirlos al canal de youtube de la materia (el cual se abrió para la ocasión). Un docente, que tenía experiencia en el uso de recursos audiovisuales fue el responsable de la tarea de edición, manejo y análisis de datos del canal del canal.

Se han subido al canal 12 videos teóricos, los cuales tienen una duración dispar. Los mismos fueron grabados durante el desarrollo del curso y fueron presentados a los estudiantes a medida que se iban desarrollando los diversos temas. También se grabó otro video, con el objetivo de dar a conocer al estudiante el formato de las pruebas virtuales -que pasaron de dos parciales a cuatro pruebas que incluían 2 quick test- y cómo usar la plataforma para resolverla. Asimismo se diseñó un simulacro de prueba, que se puso a disposición de los alumnos. El video mostraba también cómo acceder al simulacro pasando luego a explicitar los aspectos técnicos que deberían considerar al responder y entregar virtualmente la prueba.

Cabe destacar que coincidentemente con el segundo parcial que indicaba la finalización del curso, hubo período de examen. El público del examen está compuesto por estudiantes que no superaron el primer parcial, alumnos de generaciones anteriores y estudiantes libres.

A efectos de alcanzar el objetivo de la investigación, la metodología de análisis se basó en la herramienta Youtube Studio, de libre acceso, que brinda estadísticas sobre el comportamiento de los usuarios del canal.

### 4) RESULTADOS OBTENIDOS

El canal de youtube "UA Costos y Control de Gestión" en el período comprendido entre su creación y hasta el 22 de julio, fecha del segundo parcial y examen ha tenido 16.340[9] visualizaciones, provenientes de al menos 1486 espectadores. El número de espectadores surge de la audiencia que ha tenido el video más reproducido, que fue el de explicación del simulacro.

Los datos indican que de los visitantes del canal registrados en youtube, el 99% tienen una edad entre 18 y 34 años, y el público está compuesto por un 52,7% de mujeres y 47,3% hombres.

Con respecto al origen de las visualizaciones, la fuente principal proviene del sitio de la materia en la plataforma Moodle, desde donde se realizó el 85,4% de los accesos a las visualizaciones.

En cuanto al número de visitantes del canal y el público objetivo, se sabe que rindieron el primer parcial 970 alumnos. Esta cifra es levemente superior al promedio de los últimos 3 años y muestra una estabilidad[10] que permite afirman que no hubo efectos negativos con respecto a la posible deserción de estudiantes provocado por el cambio metodológico. Desde la creación del canal hasta el 22 de mayo de 2.000, fecha de la primera prueba parcial, se alojaban en dicho canal 6 videos. El primero contenía la explicación del simulacro de prueba ya mencionado y los otros 5 correspondían a contenidos teóricos. Estos últimos tuvieron una audiencia comprendida entre los 772 y 987 espectadores, lo que implica que el acceso al recurso audiovisual se encuentra entre el 80 y el 100% del público objetivo.

---

[9] Cabe destacar que algunos videos fueron vistos vía BBB, y sus visualizaciones se adicionaron al canal.

[10] 969, 970 y 968 alumnos presentados en los años 2017, 2018 y 2019 respectivamente.

Se observó que las visualizaciones no son constantes en el tiempo. Se produce un primer pico de visualizaciones cuando se crea el canal de youtube. Luego la concentración corresponde a las fechas previas a las pruebas parciales y examen (22/05 y 22/07), y también un aumento no tan considerable pero destacable en las visitas previas a los *quick test* que se realizaron el 06/06 y el 23/06.

Gráfico 1: Cadencia de Visualizaciones

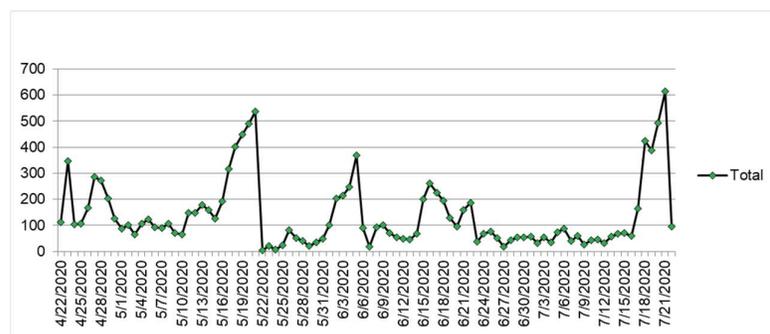

Fuente: Elaboración Propia

A efectos de una mejor visualización del incremento en las visualizaciones previo a una prueba, se muestra la cadencia de los primeros 22 días del mes de julio, mes en que no hubo *quick test* y con fecha 22 se realizó el segundo parcial y el examen

Grafico 2: Cadencia de Visualizaciones período 1 al 22 de julio

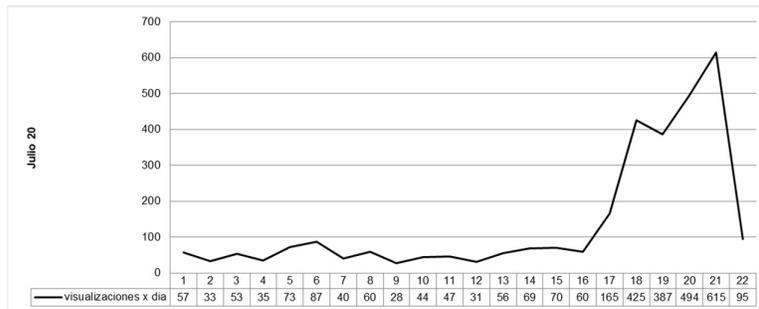

Fuente: Elaboración Propia

Respecto al tiempo de visualización en el total del período, se analizará por separado el video explicativo del simulacro de prueba, de los 12 videos de contenido teórico. Estos últimos tuvieron una duración media de 23 minutos con 53 segundos.

El tiempo promedio de visualización es de 10 minutos 4 segundos. Individualmente, el video que alcanzó el mayor tiempo de visualización promedio fue de 17 minutos 26 segundos, y el menor de 2 minutos 21 segundos

Aunque las curvas de duración del video y de tiempo de visualización promedio tienen un comportamiento similar, la dispersión entre los valores y la media es menos significativa en el grafico de tiempo de visualización, tal como puede observarse en el gráfico siguiente.

Grafico 3: Duración del Video y Tiempo de visualización unitario

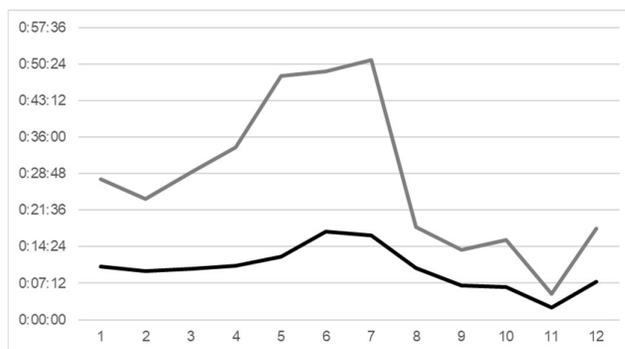

Fuente: Elaboración Propia

Dado que cada espectador realiza más de una visualización (en promedio cada video es visto 1,82 veces con una mínima de 1,34 y una máxima de 2,06), se procedió a analizar el tiempo total que cada espectador promedio le dedicó a cada video. Los resultados pueden verse en el siguiente gráfico:

Gráfico 4: Duración del Video y Tiempo dedicado por espectador

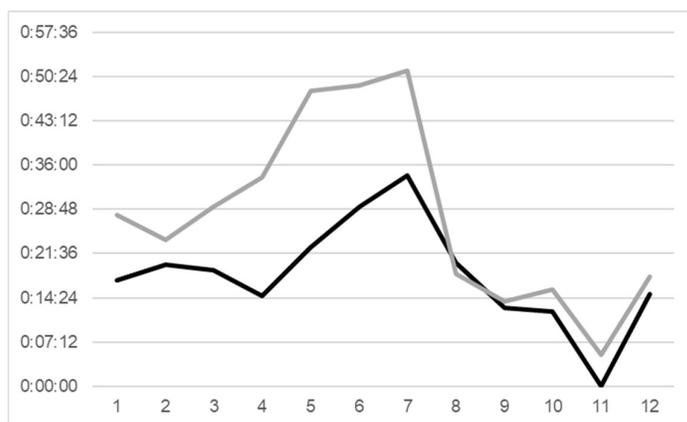

Al introducir la variable visualización/espectador al análisis de los tiempos, se puede efectuar una aproximación a la cobertura temática del video captada por el estudiante[11]. En promedio la cobertura tuvo una media de 72,5% , observándose que los videos de corta duración fueron los que captaron una mayor cobertura.

Los 12 videos recientemente analizados, tienen diferente duración, contenido e intervienen distintos docentes, pero todos tenían el propósito de formar al estudiante en los conceptos teóricos que dan base a la materia. Como ya fue expuesto, existió otro video, denominado explicación del simulacro de prueba, cuyo objetivo fue mostrar al estudiante cómo se desarrollaría la prueba virtual y cómo acceder y completar una prueba simulada.

Dicho video fue reproducido al finalizar el período analizado por 1.486 estudiantes alcanzando las 2.334 visualizaciones cifras superan significativamente al promedio de espectadores y visualizaciones de los 12 videos teóricos (227 y 200% respectivamente)

---

[11] Sería la cobertura máxima posible, dado que el estudiante puede volver a ver los mismos contenidos.

La cadencia de visualizaciones, hasta el 22 de julio de 2.000 se muestra a continuación:

Gráfico 5 Cadencia de visualizaciones video explicativo

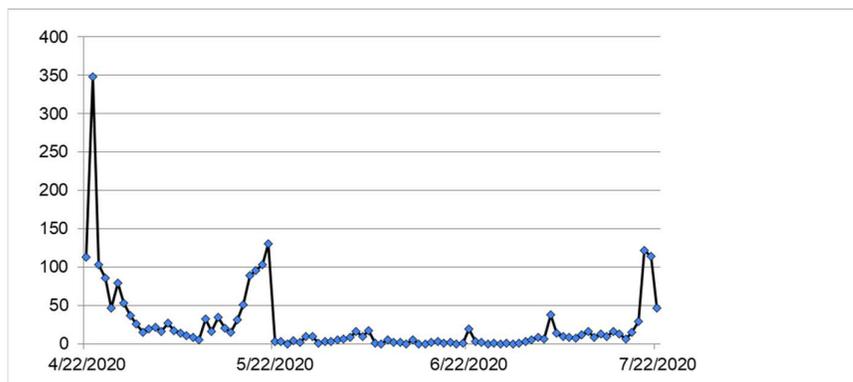

Fuente: Elaboración Propia

El gráfico muestra una acumulación de las visualizaciones al momento de la publicación del video, un pico de visualizaciones el día antes del primer parcial, y nuevamente se observa un incremento en los días antes del segundo parcial, que coincide con el examen. Esto era esperable si se considera que el público del examen está dado por estudiantes que no aprobaron el primer parcial y un nuevo público compuesto por estudiantes de generaciones anteriores y libres.

El tiempo promedio de las visualizaciones fue de 8 minutos 7 segundos, cuando la duración del video es de 27 minutos con 29 segundos. Las visualizaciones por visitante se determinaron en 1,57, por lo que el tiempo total que un visitante medio le dedicó a este video fue de 12 minutos con 45 segundos. En promedio, el tiempo dedicado por el estudiante captó a lo sumo el 46,3% del contenido, ocupando el segundo lugar en la baja captación de la muestra y el de menor relación visualizaciones/espectador. Cabe destacar que el simulacro de prueba fue realizado por 724 estudiantes que utilizaron 1.544 intentos de prueba.

## 5) CONCLUSIONES

La escasa formación pedagógica que suele ser usual entre los docentes universitarios, junto con el desconocimiento de las tecnologías a aplicar en las modalidades a distancia, posiciona a los docentes en una ubicación desfavorable dentro del TPACK. Sin embargo, la respuesta pedagógica ante la crisis sanitaria provocada por el COVID 19 dejó de manifiesto que en la materia Modelos y Sistemas de Costos se recurrieron a nuevas herramientas didácticas, a la incorporación de un CAB con buena aceptación por parte del alumnado y que no hubo deserciones de la matrícula motivada por el cambio de metodología.

La formación acerca del conocimiento tecnológico adecuado (TK) resulta imprescindible y se logró el avance de formar docentes en la programación y manejo de las pruebas virtuales así como en la elaboración de recursos didácticos audiovisuales. Con respecto a estos últimos, tuvieron una buena receptividad del alumnado con un acceso a los mismos superior al 80% Sin embargo la eficiencia del recurso audiovisual no fue máxima, ya que la duración de estos excedió lo recomendado en el marco teórico planteado. Cuando se acortó el tiempo de duración, dividiendo la temática a abordar en varios videos, se produjo una mejora en la eficacia y la eficiencia.

Con respecto al video de explicación del simulacro de prueba, la alta demanda de espectadores y visualizaciones ratifica la posición ya señalada de Toki y Caukill (2003) sobre la importancia de la implementación de simulacros y que el estudiante se familiarice con el formato de prueba. La interpretar de la poca captación del interés del público por finalizar el video, derivada probablemente a su formación como nativos digitales, deja un camino abierto a futuras investigaciones en la línea de autores como Joaquin Brunner o Carlos Tadesco quienes analizan la compleja relación entre las nuevas tecnologías aplicadas a la educación, dando marcos teóricos que sustentan las bases para una educación del futuro.